# Development of an Entropy-Based Feature Selection Method and Analysis of Online Reviews on Real Estate

Hiroki Horino[1], Hirofumi Nonaka[1], Elisa Claire Alemán Carreón[1], Toru Hiraoka[2]
[1]Department of Information and Management System Engineering, Nagaoka University of Technology, Nagaoka, Japan
(s153418@stn.nagaokaut.ac.jp, nonaka@kjs.nagaokaut.ac.jp, s153400@stn.nagaokaut.ac.jp)
[2]Department of Information Systems, University of Nagasaki, Nagasaki, Japan
(hiraoka@sun.ac.jp)

*Abstract* – **In recent years, data posted about real estate on the Internet is currently increasing. In this study, in order to analyze user needs for real estate, we focus on "Mansion Community" which is a Japanese bulletin board system (hereinafter referred to as BBS) about Japanese real estate. In our study, extraction of keywords is performed based on calculation of the entropy value of each word, and we used them as features in a machine learning classifier to analyze 6 million posts at "Mansion Community". As a result, we achieved a 0.69 F-measure and found that the customers are particularly concerned about the facility of apartment, access, and price of an apartment.**

## I. INTRODUCTION

The rapid growth of Internet applications on real estate leads to a large amount of apartment online reviews. Analyzing these reviews based on Text mining is useful for a deep understanding of consumer behavioral insights.

In recent years, Text mining is widely used for economy and management researches. Bollen et al., [1] predicted the Dow Jones Industrial Average by extracting public mood from tweets. They used the mood tracking tool GPOMS and achieved high accuracy in predicting the daily changes in the closing values of the Dow Jones Industrial Average. There are several other studies on relationship between text data and economic data. Asur et al., [2] forecasted box-office revenues for movies using Twitter data. Mao [3] used Mobile communication data to measure socio-economic indicators in Côte d'Ivoire. O'Connor et al., [4] conducted correlation analysis between consumer confidence and sentiment word frequencies in contemporaneous tweets.
Nonaka et al, [5] developed a patent scoring method based on citation networks and analyzed relation between the score and stock price.

On the other hand, some researchers attempted to analyze user needs by using Text mining. He et.al., [6] analyzed web marketing of pizza industry by text-mining tools applying to Tweets and Facebook data. Nonaka et al, [7, 8] extracted technology words and user needs from patent documents by using an entropy and grammatical pattern based method.

However, on the research fields of analyzing real estate market, many researchers use traditional economic data only [9, 10]. Hence, there are few researches using text data. Zamani et al., [11] attempted to predict real estate at the level of US country using the LDA [12] topics of tweets (not only related to real estates). This research did not focus on user needs.

In this study, in order to analyze user needs for real estate, we focus on "Mansion Community" which is Japanese BBS on Japanese real estate because detecting user reviews is easier than from SNS data. And we proposed an entropy based feature selection method and applied it to the classification of the data extracted from "Mansion Community" by using SVM.

This paper is organized as follows. In section 2, we explain about methodology. We present our experiments of system performance and analysis result in section 3. Based on this system, we conduct an evaluation including the discussions in section 4. Finally, section 5 concludes this paper.

## II. METHODOLOGY

### 2.1 Overview

Our proposed method is shown in fig 1.

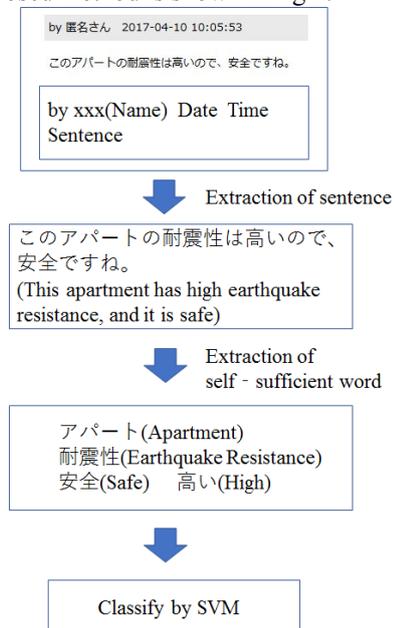

**Fig 1.** Outline of our method

### 2.2 Word Segmentation

For an analysis to be made possible for each word, we segmented the collected Japanese texts without spaces into words using the MeCab [13]. After segment the words, we extracted only a self- sufficient word.

## 2.3 Entropy Based Keyword Extraction

Feature selection of our method is based on the Shannon's entropy (hereinafter referred to entropy) value [14] of each word. According to Information Theory, entropy is the expected value of the information content in a signal [14].

Applying this knowledge to the study of words allows us to observe the probability distribution of any given word inside the corpus. For example, a word that keeps reappearing in many different documents will have a high entropy, given that predicting on which document it would appear becomes uncertain. On the contrary, a word that only was used in a single text and not in any other documents in the corpus will be perfectly predictable to only appear in that single document, bearing an entropy of zero. This concept is shown in the figure below.

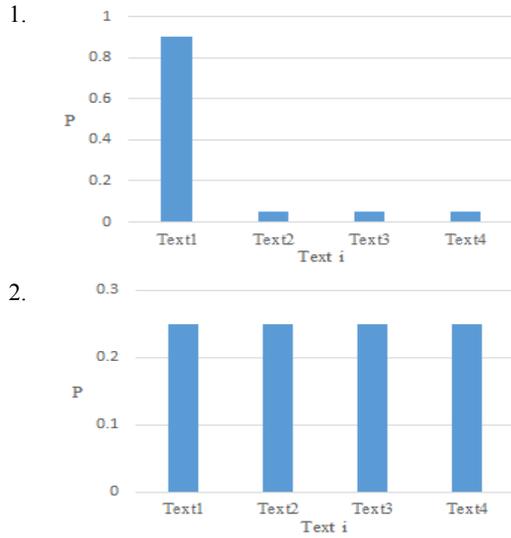

**Fig 2.** Probabilities of a word *j* being contained in a document *i*
1. Entropy close to zero, 2. High entropy

If a word has a higher entropy in positive documents than in negative documents by a factor of alpha (α), then it means its probability distribution is more spread in positive texts, meaning that it is commonly used in positive tagged documents compared to negative ones.

To calculate the entropy in a set of documents, for each word *j* that appears in each document *i*, we counted the number of times a word appears in positive comments as $N_{ijP}$, and the number of times a word appears in negative comments as $N_{ijN}$. Then, as shown in the formulas below, we calculated the probability of each word appearing in each document shown below as $P_{ijP}$ (1) and $P_{ijN}$ (2).

$$P_{ijP} = \frac{N_{ijP}}{\sum_{i=1}^{M} N_{ijP}} \quad (1)$$

$$P_{ijN} = \frac{N_{ijN}}{\sum_{i=1}^{M} N_{ijN}} \quad (2)$$

We then substitute these values in the formula that defines Shannon's Entropy. We calculated the entropy for each word *j* in relation to positive documents as $H_{Pj}$ (3), and the entropy for each word *j* in relation to negative texts as $H_{Nj}$ (5). That is, as is shown in (4) and (6), all instances of the summation when the probabilities $P_{ijP}$ or $P_{ijN}$ are zero and the logarithm of these becomes undefined are substituted as zero into (3) and (5).

$$H_{Pj} = -\sum_{i=1}^{M} \left[ P_{ijP} \log_2 \left( P_{ijP} \right) \right] \quad (3)$$

$$\lim_{P_{ijP} \to 0+} P_{ijP} \log_2(P_{ijP}) = 0 \quad (4)$$

$$H_{Nj} = -\sum_{i=1}^{M} \left[ P_{ijN} \log_2 \left( P_{ijN} \right) \right] \quad (5)$$

$$\lim_{P_{ijN} \to 0+} P_{ijN} \log_2(P_{ijN}) = 0 \quad (6)$$

After calculating the positive and negative entropies for each word, we measured their proportion using the mutually independent coefficients α for positive keywords and α' for negative keywords, for which we applied several values experimentally. A positive keyword is determined when (7) is true, and likewise, a negative keyword is determined when (8) is true.

$$H_{Pj} > \alpha H_{Nj} \quad (7)$$

$$H_{Nj} > \alpha' H_{Pj} \quad (8)$$

## 2.3 Word Analysis Using Support Vector Machine

In machine learning, Support Vector Machines are supervised learning models commonly used for statistical classification or regression [15]. Using already classified and labeled data with certain features and characteristics, an SVM learns to classify new unlabeled data by drawing the separating (p-1)-dimensional hyperplane in a p-dimensional space. Each dimensional plane is represented by one of the features that a data point holds. Then each data point holds a position in this multi-dimensional space depending on its features. The separating hyperplane and the supporting vectors divide the multi-dimensional space by minimizing the error of classification. A two-dimensional example is shown below.

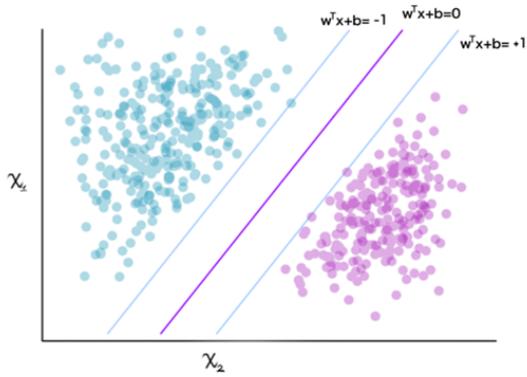

**Fig 3.** 2D example of the Linear SVM classification

The linear kernel for the SVM classification is defined by the formula (9) below. The influence that each point of the training data inputs into the vector is defined by their weight $w_n$, included in the Weight Vector $w$. The bias coefficient b determines the position of the hyperplane.

$$f(x) = w^T x + b \quad (9)$$

Then, the conditions shown in (10) are applied when classifying new data.

$$f(x) \begin{cases} \geq 0 & y_i = +1 \\ < 0 & y_i = -1 \end{cases} \quad (10)$$

Now, the initial condition is set as w = 0. Then each possible separating vector is tested, and when a classification for $x_i$ fails, the value for w is changed as follows in (11) by a value of α.

$$w \leftarrow w + \alpha sign(f(x_i))x_i \quad (11)$$

This process is repeated until all points of the training data are classified correctly. The resulting formula for classification (12) is as follows.

$$f(x) = \sum_{i=1}^{N} \alpha_i y_i (x_i^T x) + b \quad (12)$$

In Natural Language Processing, when it comes to statistically analyzing documents, each possible word in that corpus is a feature, or a dimensional plane. Then, the value of that feature will be marked as the number of times a word j is contained in a document i.

We implemented this theory in Python using the Support Vector Classifier (SVC) included in the library *scikit-learn*[1]. To vectorize the documents into word vector spaces we used the method *CountVectorizer* included in the same library. We then managed these vectors using the mathematics library *numpy*[2].

To evaluate each of our trained machines, we used the K-fold Cross Validation method, which has been proven to provide good results [16]. In each test, we calculated then the Precision, Recall, $F_1$ [17] and Accuracy values for our predictions.

## III. RESULTS

### 3.1 Experiments of system performance

Before analysis by using Support Vector Machines, we defined five topics by manually. In "Mansion Community"[3], posts are mainly mentioned about five topics. We made the training data by randomly collecting 500 posts and classified into five topics.

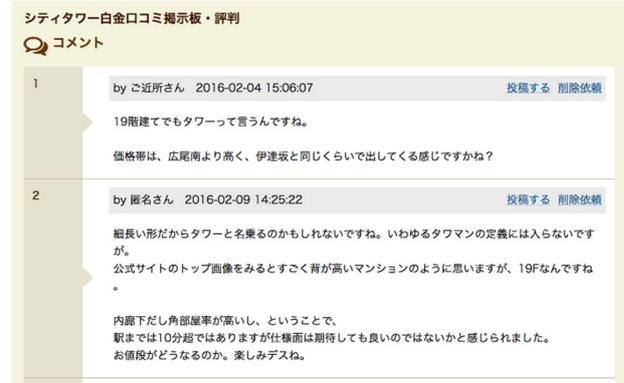

**Fig 4.** Example posts in "Mansion Community"

Table 1 shows the topic content and example keywords for each topic obtained from Entropy.

---

[1] Scikit-Learn http://scikit-learn.org/

[2] Numpy http://numpy.org/

[3] Mansion Community https://www.e-mansion.co.jp/

Table 1
Topic content and example keywords of each topic

|  | Topic content | Example keywords |
|---|---|---|
| Topic 1 | About apartment access | "駅"(*eki*; station) "近い"(*chikai*: near) |
| Topic 2 | About apartment security | "オートロック"(*ōtorokku*; automatic lock) "セキュリティ"(*sekyuriti*; security) |
| Topic 3 | About surrounding facilities | "公園"(*kōen*; park) "スーパー"(*sūpā*; supermarket) |
| Topic 4 | About facility of apartment | "エレベーター"(*erebētā*; elevator) "駐車場"(*chūshajō*; parking lot) |
| Topic5 | About contract, price of apartment | "円"(*en*; Japanese yen) "高い"(*Takai*; expensive) |

We crawled a total of 49,683 html files from "Mansion Community". "Mansion Community" is a BBS site which handle information about real estate in Japan. From these html files, we extracted a total of 6,320,631 posts.

We classified posts into one of the five topics shown on above by using Support Vector Machines. To evaluate each of our trained machines, we calculated then the Precision, Recall, $F_1$-Score.

Precision($P$) is defined as the number of true positives ($T_p$) over the number of true positives plus the number of false positives ($F_p$).

$$Precision(P) = \frac{T_P}{T_P + F_P} \qquad (13)$$

Recall($R$) is defined as the number of true positives ($T_p$) over the number of true positives plus the number of false negatives ($F_n$).

$$Recall(R) = \frac{T_P}{T_P + F_N} \qquad (14)$$

$F_1$-Score is a measure of a test's accuracy, and it considers both the Precision and the Recall of the test to compute the score. $F_1$-Score is the harmonic mean of Precision and Recall, and it is can be calculated from the following formula.

$$F_1 - Score = \frac{2 * Precision * Recall}{Precision + Recall} \qquad (15)$$

Table 2 shows the results of the Precision, Recall, $F_1$-Score in each topic.

Table 2
Results of the Precision, Recall, $F_1$-Score in each topic

|  | Precision | Recall | $F_1$-Score |
|---|---|---|---|
| Topic 1 | 0.95 | 0.65 | 0.77 |
| Topic 2 | 1.00 | 0.57 | 0.73 |
| Topic 3 | 1.00 | 0.43 | 0.60 |
| Topic 4 | 0.92 | 0.43 | 0.59 |
| Topic 5 | 0.96 | 0.63 | 0.76 |

**3.2 Result of Analysis**

We crawled a total of 49,683 html files from "Mansion Community"1 for data analysis of user needs. From these html files, we extracted a total of 6,320,631 posts. And then we applied SVM to classify 5 topics. Table 3 shows the results of number of user comment on each topic.

Table 3
Results of number of Positives

|  | Number of User Comments |
|---|---|
| Topic 1 | 928,885 |
| Topic 2 | 151,214 |
| Topic 3 | 558,947 |
| Topic 4 | 1,165,116 |
| Topic 5 | 896,317 |

Most mentioned topic was topic4, which mean facility of apartment and next was topic1, which mean access of apartment. Whereas, most less mentioned topic was topic 2, which mean security of apartment.

IV. DISCUSSION

First, we will focus on the evaluation of the system performance. Our method achieved 0.69 of the average $F_1$ of five topics and 0.97 of the average precision which can be considered as high. However, there is still room for improvement of recall. Therefore, in future work, we will optimize the coefficients of (7) and (8) which controls varieties of feature words in each topic.

Second, we will discuss the results of data analysis. Table 3 showed the most mentioned topic was about "facility of apartment". In this topic, earthquake related word was often used such as "このマンションは耐震性が高く、安全ですね。" (Kono manshon wa taishin-sei ga takaku, anzen desu ne.; This apartment has high earthquake resistance, and it is safe.). In addition, there were keywords related to earthquakes, such as "免震構造" (Menshin kōzō; quake-absorbing structure), and "地震" (Jishin; earthquake) in topic 4. Earthquakes are frequent in Japan; therefore, many Japanese customers are concerned about the earthquake resistance of an apartment.

V. CONCLUSION AND FUTURE WORKS

In our study, with the objective to understand what Japanese customers of Japanese real estate demand, we extracted keywords from a Japanese BBS site about Japanese real estate.

Regarding the needs of Japanese customers of Japanese real estate, we found that the customers are concerned about the access, security, surrounding facilities, price and facilities of an apartment. The most mentioned topic was facility of apartment, which include keywords related to earthquakes. As we mentioned in section IV, the frequent earthquakes in Japan can be thought to have had influenced this result.

For our future work, we need to improve the average $F_1$-Score of our five topics. To achieve this, we need to increase training data and eliminate number of training data between topics. Topic 2 had less training data than other topics and topic 1 has the most training data, nearly three times topic 2 and this might cause affected for keyword of topic 2.

REFERENCES

[1] Johan Bollen, Huina Mao, Xiaojun Zeng. Twitter mood predicts the stock market. Journal of Computational Science, 2(1), pp.1–8, 2011.
[2] Sitaram Asur, Bernardo A. Huberman. Predicting the future with social media. Proc. IEEE/WIC/ACM International Conference on Web Intelligence and Intelligent Agent Technology (WI-IAT) 2010, volume 1, pp.492–499.2010.
[3] Huina Mao, Xin Shuai, Yong-Yeol Ahn, Johan Bollen, Quantifying socio-economic indicators in developing countries from mobile phone communication data: applications to Côte d'Ivoire. EPJ Data Science, 4(1), 15, 2015.
[4] Brendan O'Connor, Ramnath Balasubramanyan, Bryan R. Routledge, Noah A. Smith..From tweets to polls: Linking text sentiment to public opinion time series. Proc. ICWSM, 11(122-129), pp.1–2, 2010.
[5] Hirofumi Nonaka, Daiki Kubo, Toru Kimura, Takahisa Ota, Shigeru Masuyama, Correlation analysis between financial data and patent score based on HITS algorithm, Proc. IEEE Technology Management Conference (ITMC), pp. 1-4, 2014.
[6] Wu Hea, Shenghua Zhab, Ling Lia, Social media competitive analysis and text mining: A case study in the pizza industry, International Journal of Information Management Volume 33, Issue 3, pp. 464–472, 2013.
[7] Hirofumi Nonaka, Akio Kobayahi, Hiroki Sakaji, Yusuke Suzuki, Hiroyuki Sakai, Shigeru Masuyama, Extraction of the Effect and the Technology Terms from a Patent Document, Proc. 40th International Computers and Industrial Engineering (CIE), 2010.
[8] Hirofumi Nonaka, Akio Kobayahi, Hiroki Sakaji, Yusuke Suzuki, Hiroyuki Sakai, Shigeru Masuyama, Extraction of the Effect and the Technology Terms from a Patent Document, Journal of Japan Industrial Management Association, vol. 63, no. 2E, pp.105-111, 2012.
[9] Francois Ortalo-Magne, Sven Rady. Housing market dynamics: On the contribution of income shocks and credit constraints. The Review of Economic Studies, 73(2):459–485, 2006.
[10] Badi H. Baltagi , Jing Li, Cointegration of matched home purchases and rental price indexes — Evidence from Singapore, Regional Science and Urban Economics, 55, pp.80–88, 2015.
[11] Mohammadzaman Zamani, Hansen Andrew Schwartz, Using Twitter Language to Predict the Real Estate Market, EACL 2017, pp. 28-33, 2017
[12] Blei, D.M., Ng, A.Y, Jordan, M.I. Latent Dirichlet Allocation, Journal of Machine Learning Research, pp. 933-1022, 2003.
[13] Taku Kudo, Kaoru Yamamoto, Yuji Matsumoto, Applying Conditional Random Fields to Japanese Morphological Analysis, Proceedings of the 2004 Conference on Empirical Methods in Natural Language Processing (EMNLP-2004), pp.230-237, 2004.
[14] Shannon E Claude, A Mathematical Theory of Communication, Bell System Technical Jour-nal. Vol. 27, No. 3, pp. 379 – 423, 1948.
[15] Cortes, C., Vapnik, V.: Support-vector networks. In: Machine Learning. Vol. 20, No. 3, pp. 273-297, 1995.
[16] Kohavi, R.: A study of cross-validation and bootstrap for accuracy estimation and model selection, Ijcai. Vol. 14. No. 2,1995.
[17] Powers, D.: Evaluation: From Precision, Recall and F-Measure to ROC, Informedness, Markedness & Correlation. Journal of Machine Learning Technologies, Vol. 2, No. 1, pp. 37-63, 2011.